\documentstyle[preprint,aps]{revtex}
\begin{document}
\tightenlines 
\title{
Why odd-space and odd-time dimensions in even-dimensional
spaces?\footnote{NBI-HE-00-26}}
\author{ N. Manko\v c Bor\v stnik}
\address{ Department of Physics, University of
Ljubljana, Jadranska 19,\\
and J. Stefan Institute, Jamova 39,\\
Ljubljana, 1111, \\
and Primorska Institute for Natural Sciences and Technology,\\
C. Mare\v zganskega upora 2, Koper 6000, Slovenia}
\author{ H. B. Nielsen}
\address{Department of Physics, Niels Bohr Institute,
Blegdamsvej 17,\\
Copenhagen, DK-2100,\\
and TH Division, CERN,
CH-1211 \\
Geneva 23, Switzerland }
 \maketitle
 \begin{abstract}
We are answering  the question why
$4$-dimensional space has the
metric $1+3$ by making a general argument from
 a certain type of equations of motion linear in momentum
for any spin (except spin zero) in any even dimension d. All
known free
equations for non-zero spin  for massless fields
belong to this type of equations.
Requiring Hermiticity\footnote{This is a generalization of an
earlier work which shows that without  assuming the Lorentz
invariance -which in the present work {\it is} assumed- the Weyl
equation follows using Hermiticity.} of the equations
of motion operator
as well as irreducibility with respect to the Lorentz group
representation, we prove that only metrics with the signature
corresponding to
 $ q$ time + $(d - q)$ space dimensions with $q$
{\em being odd} exist.
Correspondingly, in four
dimensional space, Nature could only make the realization of
$1+3$-dimensional space.
\end{abstract}
\pacs{
}
 
 
{\it Introduction:}

One of the most exciting open questions in physics and
cosmology\cite{Hawking}
is why the metric of our World is the Minkowski one. With the
metric of only the Euclidean signature, for example, the World
would have almost no dynamics (classically) for $p_1^2 + p_2^2 + p_3^2 +
p_4^2 = m^2 $, and 
massless particles like photons, gluons and gravitons would not
even exist (onshell) (provided that $p^a,\; a \in \{1,2,3,4 \}$ are real
quantities, which of course is understood, since otherwise the
signature
loses sense). Concerning the problem of dynamics, two space and
two time signatures would work but other problems would occur
\cite{Tegmark},
like the problem of causality.

The argument presented here comes entirely from considering the
{\em internal space},i.e. spin,
degrees of freedom.
 
Also with internal degrees of freedom arguments, S. E. Rugh and
one of
us\cite{npbps29} argue that, if one requires 
 the equation operator
to be Hermitian, then although  the Lorentz invariance is
assumed to be broken at the outset, an equation which
is Lorentz invariant and  can be interpreted as the
Weyl equation follows as an especially stable
possibility, while the space-time is (1+3)-dimensional
\cite{ORD},\cite{Why3plus1},\cite{FroggattNielsenbook}.
 
J. Greensite\cite{Greensite92}\cite{Greensite93} has argued for
the signature of (1+3) by assuming that the Lorentz symmetry is
a dynamical quantity. Hawking\cite{Hawking}, on the other hand 
has pointed to
the Wheeler-De-Witt equation, in which the signature cannot even
be seen, and so, he says, it
should not matter whether the metric is either Euclidean or
Minkowskian. Tegmark\cite{Tegmark}
has given antropical principle arguments in a spirit of
random dynamics, very much like that by one of
us\cite{ORD}\cite{FroggattNielsenbook}.
Weinberg has an argument for one-time signature in string
theory\cite{weinberg}.
Penrose and Rindler have some remark on the special properties
of the experimental signature and dimension for the Weyl tensor
\cite{Penrose}.

The present work is based on considering free equations of motion
for an arbitrary spin and dimension and signature. We shall use a
special form of such equations put forward by one of us,
who\cite{norma92,norma93,norma94,norma95,norma99} has
proposed the approach in
which all the
internal degrees of freedom are described as the dynamics in the
space of anticommuting coordinates,
that is, the Lorentz symmetry in the
internal space of anticommuting coordinates manifests in the
four-dimensional space as spins
and charges for either fermions or bosons. This approach 
offers an easy and elegant way to define the  equations of
motion for
 fermions or bosons to any dimension d (either even or odd).
We have made use in this paper of some of the results presented in the
references\cite{norma93,normapika,normaholger,nf,bojannorma}.

 
 {\em In this paper, we are presenting a general proof  that under
certain
 assumptions, for any even-dimensional space, the
number of, say, time dimensions must be odd.}
We take the starting point of writing a rather general
form of  equations of motion
\begin{eqnarray}
{\bf B {\cal P}}^{a_0}\; \psi&=&0,
\nonumber
\\
{\rm where} \quad \quad \quad \quad
\nonumber
\\
{\bf {\cal P}}^{a_0}&=& p^{a_0}\; +\; i\; \alpha \;{\bf
S}^{a_0i}\;p^i \eta^{ii}, \quad i \ne
 a_{a_0},
\label{maineq}
\end{eqnarray}
which is obeyed by all known free fields (the Weyl fields, the Yang-Mills
fields), except a scalar one, like the Higgs one
(if it exists). Here $p^a$ is the d-momentum, ${\bf
S}^{ab}$ are the Lorentz
generators in internal space; it is the Lorentz generators acting on
the spin states. The operator ${\bf B}$ is a spin-space matrix,
which because
of our assumption about ``irreducibility'' of the representation is
a function of the ${\bf S}^{ab}$'s, but not (necessarily)
Lorentz invariant.
The constant $\alpha $ is equal to the inverse of the
(maximal) helicity of the field  and is for fermions equal to
two,  for vectors, equal to one, and so on. In fact equation (\ref{maineq})
is obtainable by rewriting
equations of motion like the Weyl equations and the Maxwell equations.
We  also take
gravity as a free field and only care for it as gravitons.
 
 
The general proof follows if the  following basic assumptions are
fulfilled:
 
\noindent
1. {\em Equations of motion are of the form} (\ref{maineq}).
(This assumption implies that the equations of motion are linear
in the d-momentum and thus that we consider massless particles.
It also implies that the equation content is Lorentz invariant,
but that the form of it is not manifestly Lorentz invariant.)
 
\noindent
2. {\em The equations of motion operator is  Hermitian}.
 
\noindent
3. {\em The equations of motion operator operates only inside an
irreducible representation with respect to the Lorentz group}.
 
We find the support for the above  assumptions in the Standard Model
of the electroweak interaction, which supposes four-dimensional
space-time. We shall argue for them in a longer article\cite{hnoddmet}.
 
 

{\it The internal  Lorentz  symmetry: }
 
The generators of the Lorentz transformations in the internal space
${\bf S}^{ab}$ fulfil the commutation relations
\begin{equation}
[{\bf S}^{ab},{\bf S}^{cd}] = i(\eta^{ad} {\bf S}^{bc}
+\eta^{bc} {\bf S}^{ad}-
\eta^{ac} {\bf S}^{bd}-\eta^{bd} {\bf S}^{ac}).
\label{sab}
\end{equation}
We recognize the generators ${\bf S}^{ab}$ to be of the
spinorial character $ S^{ab}$, if they
fulfil
the following relations
\begin{equation}
\{ S^{ab}, S^{ac}\} = \frac{1}{2}\;\eta^{aa}\;\eta^{bc},\quad {\rm no \;
 summation\; over\; a}.
\label{sabs}
\end{equation}
With the appropriate choice of the inner
product\cite{norma92,norma93,norma99,normaholger,normapika} we
can make a definition of
the Hermiticity of the spin-space Lorentz generators ${\bf
S}^{ab}$ as follows
\begin{equation}
({\bf S}^{ab})^{+} = \eta^{aa} \; \eta^{bb} \;{\bf S}^{ab},
\label{sabh}
\end{equation}
where ${ }^+$ stays for Hermitian conjugation. (This definition
agrees with the
Hermeticity properties of  $S^{ab} = - \frac{i}{4}[\gamma^a,
\gamma^b]$,
if expressed in terms of  the $\gamma^a$ matrices: $(S^{ab})^+ =
\eta^{aa}
\eta^{bb} S^{ab}$.) We shall further comment on the inner product
and the Hermiticity conditions later.
 
For  our proof  the
 expression for the Casimir for the Lorentz group,
 operating  only in internal space
\begin{eqnarray}
{\bf \Gamma}^{(int)} = (-i)^{n+1} \frac{(2i)^{n}}{(2n)!} \;
\epsilon_{a_1 a_2 a_3
a_5,..., a_{2n-1}a_{2n}} {\bf S}^{a_1a_2}{\bf S}^{a_3a_4}\cdots
{\bf S}^{a_{2n-1}a_{2n}}
\label{gammas}
\end{eqnarray}
will be needed.
Taking into account the Hermiticity properties of the generators
${\bf S}^{ab}$
from Eq.(\ref{sabh}), one finds that
\begin{equation}
({\bf \Gamma}^{(int)})^{+} = (-)\; ( \prod_b \; \eta^{bb})\;
{\bf \Gamma}^{(int)};  \;\;b\in\{1,2,\cdots,2n \}.
\label{gammaher}
\end{equation}
${\bf \Gamma}^{(int)}$ defines left (${\bf \Gamma} = -1$) and
right (${\bf \Gamma} = 1$)
handed representations.

{\it Higher spins:}

All the definitions presented above (except Eq.(\ref{sabs}),
which is only valid for spinors) are valid for any spin in any
dimensional space-time.
To treat higher spins we use the Bargmann and Wigner
construction\cite{bargmann,lurie}, making higher spin fields out
of spinorial fields by constructing the fields with many
spinorial indices of left - $\alpha,
\beta,\gamma, \dots$
- and right - $\stackrel{.}{\alpha },\stackrel{.}{\beta}, \stackrel{.}
{\gamma}, \dots$
 handedness  and generalize it in higher dimensions
( we allow besides the totally symmetrized case, which is the only 
case needed in
four-dimensions) to the antisymmetrized case and to all possible mixed
symmetrized
cases belonging to various Young tableaux $Y$ and $\hat{Y}$ for
left- and right-handed representations, respectively
\begin{eqnarray}
\psi^{(Y\hat{Y})}_{ \alpha_1 \alpha_2 \cdots \alpha_n
\stackrel{.}{\alpha_1}
                    \stackrel{.}{\alpha_2 } \cdots
\stackrel{.}{\alpha_{m}} }
= {\cal N} \; \sum_{P,\hat{P}} \;
sign^{(Y)} (P) \; sign^{ (\hat{Y}) } (\hat{P})
               \; \psi_{ \alpha_{P(1)} \alpha_{P(2)} \cdots
\alpha_{P(n)}
           \stackrel{.}{\alpha_{\hat{P}(1)}}
           \stackrel{.}{\alpha_{\hat{P}(2)}} \cdots \stackrel{.}{
\alpha_{\hat{P}(m)} }},
\label{bgpsi}
\end{eqnarray}
where $P$ and $\hat{P}$ run over the permutations, determined by the
Young tableaux $Y$ and $\hat{Y}$, respectively and the two
$sign^{(\;\;)}$ are
determined by the two Young tableaux $Y$ and $\hat{Y}$,
respectively, so that we are projecting out
a particular irreducible representation of the Lorentz group.
${\cal N}$ takes care of the normalization of the spin field.
The generators of the Lorentz transformation for any
representation
${\bf S}^{ab}$
 are constructed using the spinorial generators $S^{ab}$
(Eq.(\ref{sabs}))
\begin{equation}
{\bf S}^{ab} = \sum^{n+m}{}_{k=1} \; I_{(1)} \otimes I_{(2)} \otimes
\cdots\otimes  S^{ab}{}_{(k)}\otimes I_{(k+1)} \cdots \otimes I_{(m+n)},
\label{sabg}
\end{equation}
where $\otimes$ means the direct product and $I_{(k)}$ stays for
the unit matrix acting on the index $\alpha_k$ or
$\stackrel{.}{\alpha}_k$ and so do
$S^{ab}{}_{(k)}$. It is easily shown that ${\bf S^{ab}}$ obey the
algebra of
the Lorentz group (Eq.(\ref{sab})).
 
 

{\it Our  theorem giving an odd number of time-dimensions in
even-dimensional
space-time:}

Theorem: Assuming for any {\em irreducible representation} (for
any spin) the equations of motion (Eq.(\ref{maineq}))
\begin{equation}
{\bf B} {\bf {\cal P}}^{a_0} \psi = {\bf B} (p^{a_0} + i
\alpha {\bf S}^{a_0 i} p^i \; \eta^{ii}) \psi = 0
\label{maineq2}
\end{equation}
and letting the signature of space-time be
defined so that the bilinear form in $p^{a_0}$
\begin{equation}
\psi^+ \; (p^{a_0} - i \alpha {\bf S}^{a_0 i}\; p^i \;
\eta^{ii}){\bf {\cal
P}}^{a_0}\; \psi,
\label{kgl}
\end{equation}
has the signature of the Klein-Gordon equation for the
space-time signature in question, it follows that
the time-dimension $q $ is odd, provided that the equations of
motion  operator is Hermitian
\begin{equation}
{\bf B {\cal P}}^{a_0}\; = \; ({\bf B {\cal P}}^{a_0})^+.
\label{hermiticity}
\end{equation}
The statement above the bilinear form should even include the fact that
the positive and the negative definite subspaces of dimensions
say $q$ and $d-q $ can be chosen independently of the internal
space state $|\psi>$.

We shall first give a proof of the theorem for the spinorial
case, since for this case it is simple and transparent.

{\it Proof for the Weyl equation case:}

We recognize that the operator in Eq.(\ref{kgl}) in the Weyl
case equals  the equations of motion operator of the
Klein-Gordon equation $p^a p^b \eta_{ab}$,
provided that $\alpha = 2$, which is a real number  and that
Eq.(\ref{sabs}), which defines the anticommutation relations for
spinors, is taken into account.
When taking into account also the Hermiticity condition
(Eq.(\ref{hermiticity})) we have
\begin{equation}
B = B^+, \quad \quad -( B S^{a_0 i})^+ =  B S^{a_0 i}.
\label{proof1}
\end{equation}
It follows that
\begin{equation}
 B\;  S^{a_0 i} \; +  S^{a_0 i} \; \eta^{a_0
a_0}\; \eta^{ii}\; B \; = \; 0,
\label{bsf}
\end{equation}
which means that $B$ either commutes  or anticommutes with $ S^{a_0
i}$,
depending on whether $a_0$ and $i$ have the same or the opposite
signature.

Using Jacobi identities $[A,[B,C]] + [C,[A,B]] + [B,[C,A]] =0$ and
  $[A,\{B,C\}] + [C,\{A,B\}] + [B,\{C,A\}] =0$
we derive the following generalization of Eq.(\ref{bsf})
\begin{eqnarray}
[B, S^{ab}] &=& 0, \quad {\rm for } \; (-1)^{ \delta_{a_0 a} +
\delta_{a_0 b} }
 \eta^{aa} \eta^{bb} = 1
\nonumber\\
\{B, S^{ab}\} &=& 0, \quad {\rm for }\; \eta^{aa} \eta^{bb}\;
(-1)^{\delta_{a_0 a} + \delta_{a_0 b}} = -1.
\label{bsfg}
\end{eqnarray}
Let us now deduce the commutation versus anticommutation relations of
$B$ and
$ \Gamma^{(int)}$, the Casimir of the Lorentz group,
(Eq.(\ref{gammas})).
Taking into account  Eq.(\ref{bsfg})
we find
\begin{eqnarray}
[B, \Gamma^{(int)}] &=& 0, \quad {\rm for} \quad (-1)\; \prod_b
\eta^{bb}
= 1
\nonumber\\
\{B, \Gamma^{(int)}\} &=& 0, \quad {\rm for} \quad (-1)\; \prod_b
\eta^{bb}=- 1.
\label{gb}
\end{eqnarray}
Since $B$ was assumed to depend only on $ S^{ab}$, which means
that the equations of motion operator $B {\cal P}^{a_0}$
operates within only the irreducible representation, then $B$
should commute with $ \Gamma^{(int)}$.
Unless $ \Gamma^{(int)}$ = 0, which can cause problems,
we can now conclude from Eq.(\ref{gb}) that
\begin{equation}
 (-1)\; \prod_b \eta^{bb}
= 1,
\label{fpg}
\end{equation}
{\em which can only be true if the number of time coordinates and the
 number of space coordinates are odd}.

This finishes the proof for fermions, that is for the Weyl case.
Due to the generalized Bargmann-Wigner proposal for the description
of any spin field in d dimensions   out of
the Weyl spinors (Eq.(\ref{bgpsi})), we would intuitively
conclude that since $\alpha $ is real for
the Weyl equations of motion operator and since ${\bf S}^{ab}$
is a linear composition of $S^{ab}{}_{(k)}$'s, which each act on
different spinor indices, $\alpha$ should be real
for any spin equations of motion operator and accordingly
Eqs.(\ref{proof1}, \ref{bsf},\ref{bsfg},\ref{gb}) as well as the
equation (\ref{fpg}) should be valid for any spin in any
dimension.

The general proof, which we present below,  ensures that
$\alpha $ of Eq.(\ref{maineq}) is real for any spin.

{\it Proof for the general case:}

In order to perform the general proof of our signature theorem we
need a lemma, that is, an extension of a well-known theorem about
representations of compact groups to the non-compact groups, which
 tells us that any representation of a compact group
can be considered unitary with respect to an appropriate inner
product. The latter is constructed by averaging or integrating
over an arbitrary measure.

We present a slightly extended theorem as a lemma formulated for
the Lorentz group SO(q,d-q) with $q$ time and $(d-q)$ space
dimensions:

{\it Lemma:}

Let ${\bf S}^{ab} $ make up a (finite dimensional) representation
of the Lorentz group Lie algebra (Eq.(\ref{sab})). Then there
exists an operator ${\bf V}$ such that
\begin{equation}
({\bf S}^{ab})^+ = \eta^{aa}\; \eta^{bb}\; {\bf V} {\bf S}^{ab}
{\bf V}^{-1}.
\label{res}
\end{equation}

{\it Proof of lemma:} For real $\omega_{ab}$ the Lie algebra
consisiting of
elements of the form $\omega_{ab}{\bf S}^{ab}$ is the Lie
algebra for the group $SO(q,d-q)$, but if we instead let
\begin{eqnarray}
{\rm for} \; \eta^{aa}\;\eta^{bb} &= &+1 \; :\quad \omega_{ab}\; {\rm
real}\\
{\rm for} \; \eta^{aa}\; \eta^{bb} &= &-1 \; :\quad \omega_{ab}\;
{\rm purely \;  imaginary},
\end{eqnarray}
then the group generated will be the compact group $SO(d)$ instead.
In other words, for a choice of the square roots, the Lie
algebra consisting of the elements
\begin{equation}
\hat{\omega}_{ab} \; \sqrt{\eta^{aa}\; \eta^{bb}} \; {\bf S}^{ab}
\end{equation}
forms a Lie algebra
representation for the compact group $SO(d)$ when the
$\hat{\omega}_{ab}$ run
through real values.
On the compact group $SO(d)$ we can apply the idea of averaging
over the Haar measure an arbitrary inner product which, for
instance, may be the
one represented by the unit matrix ${\bf 1}$; that is to say, we
construct the ``averaged'' inner product as expressed by a matrix:
\begin{equation}
{\bf K}: =\; \int \; d^{Haar}g\;
(e^{i\hat{\omega}_{ab}(g)\sqrt{\eta^{aa}\eta^{bb}}
{\bf S}^{ab}})^+\;
{\bf 1}\;
e^{i\hat{\omega}_{ab}(g)\sqrt{\eta^{aa}\eta^{bb}}{\bf
S}^{ab}}.
\end{equation}
We have now ensured that the generators of the compact group
are Hermitian with respect to the inner product defined by ${\bf K}$,
becuase ${\bf K}$ constructed as an average over the whole
compact group
must be left invariant under similarity transformations with
representatives
$e^{i\hat{\omega}_{ab}{\bf S}^{ab}}$ of elements of this
compact group.  This in turn implies Hermiticity of the
generators for this group with respect to ${\bf K}$
being used as the inner product, i.e.
\begin{equation}
(e^{i\hat{\omega}_{ab}\sqrt{\eta^{aa}\eta^{bb}}{\bf
S}^{ab}})^+\;{\bf K}\;
e^{i\hat{\omega}_{ab}\sqrt{\eta^{aa}\eta^{bb}}{\bf S}^{ab}}\;
=\; {\bf K}
\end{equation}
or equivalently
\begin{equation}
{\bf K}\; \sqrt{\eta^{aa}\eta^{bb}}{\bf S}^{ab}
-(\sqrt{\eta^{aa}\eta^{bb}}{\bf S}^{ab})^+\; {\bf K} =0.
\label{hlem}
 \end{equation}
Since we had chosen definite values for the square roots,
we see that we get the statement of the lemma by dividing
equation (\ref{hlem})
by ${\bf K}$ from the right and putting ${\bf V}={\bf K}$.
This ends the proof of the lemma.

{\it Corollary to lemma:}

{\bf V} came out Hermitian and positive definite as an operator, and
the constructions $\sqrt{{\bf V}}\;{\bf S}^{ab}\; \sqrt{{\bf
V}^{-1}}$ will
obey Hermiticity relations of the form of  Eq.(\ref{sabh}).

The corollary is easily seen by dividing by $\sqrt{{\bf V}}$ to
the left
and $\sqrt{{\bf V}^{-1}}$ to the right  equation (\ref{res}).

Another useful corollary is the following

{\it Corollary: }

For the anticommutators of the generators of the Lorentz group
we have the operator-inequalities
\begin{eqnarray}
\sqrt{{\bf V}}(-\eta^{aa}\eta^{bb}({\bf S}^{ab})^2
-\eta^{cc}\eta^{dd}({\bf S}^{cd})^2)
\sqrt{{\bf V}^{-1}}\le
\sqrt{{\bf V}}\sqrt{\eta^{aa}\eta^{bb}\eta^{cc}\eta^{dd}}
\{ {\bf S}^{ab}, {\bf S}^{cd} \} \sqrt{{\bf V}^{-1}} \le
\nonumber
\\
\le \sqrt{{\bf V}}(\eta^{aa}\eta^{bb}({\bf S}^{ab})^2
+\eta^{cc}\eta^{dd}({\bf S}^{cd})^2)
\sqrt{{\bf V}^{-1}}
\end{eqnarray}
and thus the signs of the expression in Eq.(\ref{kgl}) for the
positive and negative
coordinate subspaces are not changed by ignoring the
anticommutator terms
in an expansion of the product of operators.

This corollary is rather easily shown by remarking that the
squares of the
Hermitian operators $ \sqrt{{\bf V}} \; (\sqrt{\eta^{aa}\; \eta^{bb}}
\;{\bf
S}^{ab} \;\pm\;
 \sqrt{\eta^{cc}\; \eta^{dd}}\; {\bf S}^{cd})\; \sqrt{{\bf
V}^{-1}}$, Hermitian
according to the first corollary, must be positive (semi)
definite as operators (matrices).

From this last corollary we easily see that in the general case
of arbitrary
representation we can ignore the commutator terms in equation
(\ref{kgl})
when investigating the condition for the signature, at least if we
consider $|\psi> $ as being an eigenstate of the Hermitian operator
$\sqrt{V}$. Then namely the extra $\sqrt{{\bf V}}$ and
$\sqrt{{\bf V}^{-1}}$ would
in fact effectively be replaced by even positive numbers and it would
not matter for conclusions on the sign of the expression in
Eq.(\ref{kgl}).

Then the requirement that the sign of Eq.(\ref{kgl}) for, say,
such $|\psi>$'s
that are eigenstates of $\sqrt{{\bf V}}$
must be the same for $p^{a_0}$ alone different from zero and for
$p^i$ alone different from zero in the case of
$\eta^{a_0a_0}\eta^{ii}=+1$, while they should lead to opposite signs
in the case $\eta^{a_0a_0}\eta^{ii}=-1$, gives us that $\alpha$
is {\it real}.

A problem may be if it is really possible that the expression of
Eq.(\ref{kgl}) for all $|\psi>$ 
 can come to show just  the signature of the group
$SO(q,d-q)$.
 However, the logic
here is that
we just have assumed that it is so, since the motivation for our
requirements
about the bilinear form is a replacement for assuming the
Klein-Gordon
equation to be valid as a consequence of our equation of motion.

Having now achieved the real $\alpha$ and the Hermiticity
properties
(\ref{res}) the condition
 for positivity of Eq.(\ref{maineq}) is
easily seen to be that
\begin{equation}
{\bf B}^+\; =\; {\bf B}
\end{equation}
and
\begin{equation}
(i\; {\bf B}\;\alpha \;{\bf S}^{a_0i})^+ =-i \eta^{a_0a_0}\;
\eta^{ii}\;  {\bf V} {\bf S}^{a_0i} {\bf V}^{-1}
\; \alpha\; {\bf  B}\; =\; i \; {\bf B} \; \alpha\; {\bf S}^{a_0i},
\end{equation}
the latter equation of which is equivalent to
that
\begin{equation}
[{\bf V}^{-1}\; {\bf B},\;{\bf S}^{a_0i}]=0 \;\;{\rm for}\;
\eta^{a_0a_0}\; \eta^{ii} = -1
\end{equation}
while
\begin{equation}
\{{\bf V}^{-1}\;B,\; {\bf S}^{a_0i}\}=0 \;\;{\rm for}\;
\eta^{a_0a_0}\; \eta^{ii} = +1.
\end{equation}
But that is to say that we reached the same commutation or
anticommutation
conditions (\ref{bsf}) as for the Weyl equation case, except that now
we have the
conditions for ${\bf V}^{-1}\; {\bf B}$ rather than simply for
${\bf B} $ itself. That
does not
matter, however, since with an irreducible representation even  ${\bf V}$
should be
expressed by the generators and must commute - not anticommute
(unless the Casimir is zero) - with the
Casimir operators, in particular with ${\bf \Gamma}^{(int)}$.

Hereby the proof ends also for the general case.

{ \it Concluding remarks:}

We have presented in this letter the general proof, valid for
all spins (except for spin zero) and all even dimensions, that
an equations of motion operator can  be
linear in the p-momentum, Hermitian and operate within
only the irreducible representations of the Lorentz group, if
space has an odd number of time dimensions and accordingly also an odd
number of space dimensions.
It is the spin degrees of freedom which determine the
signature of space-time.

The assumptions of the linearity in the p-momentum, Hermiticity
and irreducibility properties of the equations of motion
operator are rather mild assumptions. In fact we have been
inspired to make them by the Standard Electroweak Model assumptions, and 
we shall discuss the arguments favouring them in
a longer article\cite{hnoddmet}. Our general proof answers 
one of the most exciting open questions of science, namely why
our space-time signature is $3+1$, pointing out that there are
internal degrees of freedom that are responsible for the
signature.
Before concluding, we would only point out that all the known
elementary particles of the Standard Model (assuming $ d=4$)
belong (before
switching on the interactions) - not counting the Higgs as known - 
belong to one of the following groups: 

 1) Either to the spin-$\frac{1}{2}$ Weyl
particles, described by the Weyl equations of  left- and right-handed
irreducible representations (Eq.(\ref{maineq2})), with $B=1$ or
$B= \Gamma^{(int)}$ (which is known as $\gamma^5$) and $\alpha
=2$, leading to left-handed ($\Gamma = -1$) fermions of left
helicity
($\frac{\overrightarrow{p}.\overrightarrow{S}}
{|\overrightarrow{p}.\overrightarrow{S}|}=-1$) 
or to
right-handed ($\Gamma = 1$) fermions of right
helicity ($\frac{\overrightarrow{p}.\overrightarrow{S}}
{|\overrightarrow{p}.\overrightarrow{S}|}=1$) while
anti-fermions if left-handed
($\Gamma = -1$) have right-helicity
($\frac{\overrightarrow{p}.\overrightarrow{S}}
{|\overrightarrow{p}.\overrightarrow{S}|}=1$) and if
right-handed ($\Gamma = 1$) have left
helicity ($\frac{\overrightarrow{p}.\overrightarrow{S}}
{|\overrightarrow{p}.\overrightarrow{S}|}=-1$),
(The Standard Model postulates that no right-handed fermions or
anti-fermions  exist, which would carry the
weak charge. This feature of the Standard Model has appreciable 
experimental support.)

2) Or to the spin-one linear Yang-Mills equations of the Maxwell
type with $\alpha =1$ in Eq.(\ref{maineq}) and again with ${\bf
B}=1$ or ${\bf B}= {\bf \Gamma}^{(int)}$, which are less
known equations\cite{normapika,bojannorma} and which lead to
equations
$$
\left(
\matrix{
 p^0 + i\overrightarrow{p} \times & 0\cr
                                0 & p^0 - i \overrightarrow{p} \times \cr
} \right)
\left(
\matrix{
\overrightarrow{{\cal E}}_L \cr
\overrightarrow{{\cal E}}_R \cr}
\right) =0.
$$
Assuming $\overrightarrow{{\cal E}}_L = \overrightarrow{{\cal E}}
+ i \overrightarrow{{\cal B}}$ and  $  \overrightarrow{{\cal E }}_R =
\overrightarrow{{\cal E}} - i \overrightarrow{{\cal B}}$
the last equation leads to equations of motion for electric and magnetic
fields, which both obey Maxwell equations. We shall
comment on these equations  in a  longer
article\cite{hnoddmet}.
 
 

 

\section{Acknowledgement. } This work was supported by the  Ministry of
Science and Technology of Slovenia as well as by funds CHRX -
CT - 94 - 0621, INTAS 93 - 3316, INTAS - RFBR 95
- 0567, SCI-0430-C (TSTS). One of us (H.B.N.) wishes to thank many people 
for numerous 
discussions
on the related subject of why we have 3+1 dimensions,
and in most recently especially Bo Sture Skagerstam, S. E. Rugh, Y.
Takanishi, D. Bennett,  C. D. Froggatt and N. Stillits; N.M.B. would
like to
thank A. Bor\v stnik
and B. Gornik for discussions helpful for this letter.
 


\begin{thebibliography}{99}

%
\bibitem{Hawking} J. Hartle and S. W. Hawking, {\it Phys. Rev. } {\bf D
28} 2960  (1983);
R. Bousso and S. W. Hawking ``Lorentzian Condition in Quantum
Gravity'', hep-th/9807148.
%
\bibitem{Tegmark} Max Tegmark,`` On dimensionality of spacetime'',
gr-qc/9702052, or Class. Quan. Grav.14 (1997) L69-L75
%
\bibitem{npbps29} Holger Bech Nielsen and Svend E. Rugh ``Weyl Particles,
Weak
Interactions and Origin of Geometry'', {\it Nucl.Phys. B}
(Proc.Suppl.) {\bf 29
B,C}
(1992) 200-246; see especially page 219 for the signature.
%
\bibitem{ORD}Dual Strings - Section 6. Catastrophe Theory Programme
by H.B.Nielsen, in {\it Fundamentals of Quark Models }, eds. I.M.Barbour
and A.T.Davies, Scottish Universities Summer School in Physics (1976),
pp 528 -543; H.B. Nielsen in {\it Gauge Theories of the Eighties, } eds.
R.Raitio and J. Lindfors,(Springer Verlag,1983),p 288; H.B. Nielsen,
D. L. Bennett and Niels Brene, in {\it Recent Developments in Quantum Field
Theory}, eds. J. Ambj{\o}rn, B. Duurhus and J. L. Petersen ( Elsevier
Science Publishers, 1985), p253; J. Wheeler, ''Law without Law'', in
{\it Quantum Theory of Measurement,} eds. J. A. Wheeler and W. R. Zurek
(Princeton University Press, 1983), p. 182; J.A. Wheeler, ``Beyond the
Black Hole'', in {\it Some Strangeness in Proportion}, ed. H. Wolf, 
(Addison-Wesley, 1980), p. 341; R.P. Feynman, {\it The Character of Physical Law}
(MIT Press, 1965, Penguine Books 1992 );  G. F. Chew, in {\it Properties of Fundamental Interactions}, ed.
E. Zichichi, (Editrice Compositori, 1973), p. 3; C. H. Woo,
{\it ``Mission Impossible ? A look at Past Setbacks in the Search for 
Elementary
Matter and for Universal Symmetries ''}, Zentrum f{\" u}r
Interdisziplin{\" a}re Forschung, University of Bielefeld Preprint.
For a recent text on Random Dynamics see: Nicolai Stillits, Cand. scient.
thesis,
Niels Bohr Institute, Copenhagen, 1999.
 
%
\bibitem{Why3plus1} Holger Bech Nielsen and Svend E. Rugh ``Why do we
have
3+1 dimensions?'', hep-th/9407011, contribution to Wendisch-Rietz
meeting (Sept. 1992) eds. B. D\"{o}rfel and W. Wieczorek,
DESY preprint 93-013,
ISSN 0418-9833; S. Chadha and H. B. Nielsen ``Naturalness of Weyl
Equation and 3+1 Dimensionality'', in A Report on Research Activities
at the Niels Bohr Institute and Nordita (1974) p 117.
%
\bibitem{FroggattNielsenbook} Colin D. Froggatt and Holger Bech
Nielsen, Origin
of Symmtries, World Scientific Publishing Co.Pte.Ltd.,PO Box 128, Farrar
Road,
Singapore 9128, ISBN 9971-96-630-1,ISBN 9971-96-631-X (pbk)
%
\bibitem{Greensite92} J. Greensite, ``Dynamical Origin of the Lorentzian
Signature of Spacetime'',  {\it Phys. Lett.} {\bf B300} (1993) 34-37
%
\bibitem{Greensite93} A. Carlini and J. Greensite,``Why is Space-Time
Lorentzian ?'', {\it Phys. Rev. } {\bf D 49 } (1994) 866-878
%
\bibitem{weinberg}S. Weinberg, Proc.of the XXIII Int. Conf. on High Energy
Physics, Berkeley, (1986) (World Scientific,1987),p. 217.
%
\bibitem{Penrose} R. Penrose and W. Rindler, Spinors and space-time,
Cambridge University Press, 1986; the remark referred to is on page 235 between
formula (4.6.32)
and (4.6.33).
%
\bibitem{norma92}  Norma  Manko\v c Bor\v stnik,
''Spin Connection as a Superpartner of a Vielbein'',
{\it Phys. Lett.} {\bf B 292} (1992) 25-29;
''From a World-sheet
 Supersymmetry to the Dirac Equation'',
{\it Nuovo Cimento}
{\bf A  105} (1992) 1461-1471.
%
\bibitem{norma93} Norma  Manko\v c Bor\v stnik 
''Spinor and Vector Representations in Four-Dimensional
Grassmann Space'',
{\it J. Math. Phys.} {\bf 34} (1993) 3731-3745.
%
\bibitem{norma94} Norma  Manko\v c Bor\v stnik 
''Spinors, Vectors and Scalars in Grassmann Space and
Canonical Quantization for Fermions and Bosons'',
{\it Int. Jour. Mod. Phys.} {\bf A  9} (1994) 1731-1745;
''Unification of Spins and Charges in Grassmann Space'',
{\it hep-th/9408002};
''Quantum Mechanics in Grassmann Space, Supersymmetry and
Gravity'',
{\it hep-th/9406083}.
%
\bibitem{norma95} Norma  Manko\v c Bor\v stnik 
''Poincar\'e Algebra in Ordinary and Grassmann Space and
Supersymmetry'',
{\it J. Math. Phys.} {\bf 36} (1995) 1593-1601;
''Unification of Spins and Charges in Grassmann Space''
{\it Mod. Phys. Lett.} {\bf A 10}(1995) 587-595; {\it hep-th/9512050}
%
\bibitem{norma99} Norma  Manko\v c Bor\v stnik (1999)
''Unification of Spins and Charges in Grassmann Space'',
{\it hep-ph/9905357},
{\it  Proceedings of the International Workshop on ''What Comes
Beyond the Standard Model'', Bled,
Slovenia, 29 June-9 July 1998}, Ed. by N. Manko\v c Bor\v stnik,
H. B. Nielsen and  C. Froggatt,(DMFA Zalo\v zni\v stvo 1999) p. 20-29.
%
\bibitem{normapika} Norma Manko\v c Bor\v stnik and Anamarija
Bor\v stnik, ''Left- and right-handedness of fermions and
bosons'', {\it J. of Phys.{\bf G24} (1998) 963-977} (1999);
 Anamarija Bor\v stnik and Norma
Manko\v c Bor\v stnik (1999)
''Are Spins and Charges Unified? How Can One
Otherwise Understand Connection Between Handedness (Spin) and
Weak Charge?'',
{\it Proceedings of the International Workshop on
''What Comes Beyond the Standard Model, Bled,
Slovenia, 29 June-9 July 1998}
Eds. N. Manko\v c Bor\v stnik,
H. B. Nielsen and C. Froggatt, (DMFA Zalo\v zni\v stvo 1999) p. 52-57,
hep-ph/9905357, and paper in preparation.
%
\bibitem{normaholger} Norma  Manko\v c Bor\v stnik and
Holger Bech Nielsen (1999),
''Dirac-K\" ahler Approach Connected to
Quantum Mechanics in Grassmann Space'',
to appear in {\it Phys. Rev. D15};
{\it hep-th/9911032},
{\it Proceedings of the International Workshop on ''What Comes
Beyond the Standard Model'', Bled,
Slovenia, 29 June - 9 July 1998}, Eds. N. Manko\v c Bor\v stnik,
H. B. Nielsen and  C. Froggatt, (DMFA Zalo\v zni\v stvo 1999) p. 68-73;
{\it hep-ph/9905357;  hep-th/9909169}.
%
\bibitem{nf} Norma  Manko\v c Bor\v stnik and Svjetlana
Fajfer,
''Spins and Charges, the Algebra and Subalgebras of the
Group SO(1,14)'',
{\it Nuovo Cimento} {\bf B 112 } (1997) 1637-1665;
{\it hep-th/9506175}.
%
\bibitem{bojannorma} Bojan Gornik (1998), Diploma work, ''The Poincar\'
e Group and Equations of Motion for Free Particles'', Ljubljana
1998; Bojan Gornik and Norma Manko\v c Bor\v
stnik, ''Equations of Motion for Free Massive and Massless
particles and the Poincar\' e group'', paper in preparation.
%
\bibitem{hnoddmet} Norma Manko\v c Bor\v stnik and Holger Bech
Nielsen (2000), ''The internal space determines the metric of
space-time'', in preparation.
%
\bibitem{bargmann} V. Bargmann and E. P. Wigner, {\it
Proc. Nat. Sci. } (USA), {\bf 34} (1947) 211.
%
\bibitem{lurie} David Luri\' e, {\it Particles and
Fields}, John Wiley $\&$ Sons, New York 1968.
%

%
\end{thebibliography}
\end{document}